\documentclass[twocolumn,prd,superscriptaddress,nofootinbib]{revtex4-1}

\usepackage{amsfonts,amsmath,amssymb,mathrsfs}
\usepackage{hyperref}
\usepackage{color}
\usepackage{graphicx}  
\usepackage{dcolumn}   
\usepackage{bm}        
\usepackage[english]{babel}

\def\a{\alpha}

\def\r{\rho}
\def\s{\sigma}
\def\t{\tau}
\def\m{\mu}
\def\n{\nu}
\def\k{\kappa}
\def\th{\theta}
\def\g{\gamma}\def\G{\Gamma}
\def\L{t}\def\l{V}
\def\D{\Delta}
\def\la{\langle}
\def\ra{\rangle}
\def\o{\omega}\def\O{\Omega}
\def\d{\delta}
\def\p{\partial}

\def\oxthree{{\cal O}(x^3) }

\def\half{\textstyle{\frac{1}{2}}}

\def\bdoc{\begin{document}}
\def\edoc{\end{document}}
\def\bea{\begin{equation}}
\def\eea{\end{equation}}

\def\beq{\begin{eqnarray}}
\def\eeq{\end{eqnarray}}
\def\be{\begin{eqnarray}}
\def\ee{\end{eqnarray}}
\def\ben{\begin{enumerate}}
\def\een{\end{enumerate}}
\def\la{\langle}
\def\ra{\rangle}
\def\a{\alpha}
\def\g{\gamma}\def\G{\Gamma}
\def\d{\delta}\def\D{\Delta}
\def\e{\epsilon}
\def\z{\zeta}

\def\th{\theta}
\def\k{\kappa}
\def\l{t}
\def\m{\mu}
\def\n{\nu}
\def\o{\omega}
\def\p{\pi}
\def\r{\rho}
\def\s{\sigma}
\def\t{\tau}
\def\L{{\cal L}}
\def\S{\Sigma }
\def\gsim{\; \raisebox{-.8ex}{$\stackrel{\textstyle >}{\sim}$}\;}
\def\lsim{\; \raisebox{-.8ex}{$\stackrel{\textstyle <}{\sim}$}\;}
\def\gtrsim{\gsim}
\def\lessim{\lsim}
\def\loc{{\rm local}}
\def\vm{v_{\rm max}}
\def\bh{\bar{h}}
\def\del{\partial}
\def\nab{\nabla}
\def\half{{\textstyle{\frac{1}{2}}}}
\def\fourth{{\textstyle{\frac{1}{4}}}}

\def\bD{{\bf D}}
\def\bE{{\bf E}}
\def\bF{{\bf F}}
\def\bB{{\bf B}}
\def\bP{{\bf P}}
\def\bV{{\bf v}}
\def\bv{{\bf v}}
\def\bx{{\bf x}}
\def\by{{\bf y}}
\def\bz{{\bf z}}
\def\ba{{\bf a}}
\def\bd{{\bf d}}
\def\bs{{\bf s}}
\def\bn{{\bf n}}
\def\bp{{\bf p}}

\def\O{\Omega}

\def\br{{\bf r}}
\def\bnab{{\bf \nab}}

\def\tE{\tilde{E}}
\def\tL{\tilde{L}}
\def\Horava{Ho\v{r}ava }

\def\oxtwo{\mathscr{O}\left(x^2\right)}
\def\oxthree{\mathscr{O}\left(x^3\right)}
\def\oxfour{\mathscr{O}\left(x^4\right)}
\def\oxfive{\mathscr{O}\left(x^5\right)}
\def\LL{Lanczos-Lovelock}

\def\ph{\phantom}

\begin{document}
\title{
Overcharging higher curvature black holes
}
\author{Rajes Ghosh}
\email{rajes.ghosh@iitgn.ac.in }
\affiliation{Indian Institute of Technology, Gandhinagar, Gujarat 382355, India.}
\author{C. Fairoos}
\email{fairoos.c@iitgn.ac.in }
\affiliation{Indian Institute of Technology, Gandhinagar, Gujarat 382355, India.}
\author{Sudipta Sarkar}
\email{sudiptas@iitgn.ac.in}
\affiliation{Indian Institute of Technology, Gandhinagar, Gujarat 382355, India.}

\begin{abstract}
We examine the problem of overcharging extremal and near-extremal black hole solutions of Einstein-Gauss-Bonnet gravity in any dimension, generalizing the result in general relativity. We show that as in the case of general relativity, it is not possible to create a naked singularity by overcharging an extremal black hole in Einstein-Gauss-Bonnet gravity using a charged test particle. Our result suggests that the validity of the cosmic censorship hypothesis transcends beyond general relativity to well motivated higher curvature gravity.
\end{abstract}
\maketitle
\section*{Introduction}
The theory of general relativity (GR) describes the dynamics of the gravitational interaction in terms of space-time curvature. In spite of its spectacular successes, GR is an incomplete theory. This is primarily because of the existence of generic singular solutions of Einstein's field equations. The singularity theorems \cite{Penrose, Hawking, Hawking:1973uf} predict that under some reasonable assumptions, the end stage of gravitational collapse and the beginning of the Big Bang, the universe must be singular. At the singularity, the spacetime curvature diverges, and the classical description becomes ill-defined. It is widely believed that the physics at the singularity is described by some form of quantum gravity. \\

If at any time, the spacetime singularity is in causal contact with a distant observer, the relevant physics cannot be described by classical or semi-classical theories. Therefore, to have a consistent predictive ability, the future singularity must be causally disconnected from distant observers. This is indeed the case for the vacuum Schwarzschild solution, where the event horizon hides the singularity from the outside observers. But, there are also solutions of general relativity which may contain naked singularity, i.e., a singularity without an event horizon. The existence of such naked singularities in the physical universe would invalidate the applicability of classical and semi-classical physics. Hence, naked singularities are undesirable features of general relativity. These insights motivated Penrose to propose a bold conjecture; the so-called Cosmic Censorship Hypothesis (CCH) \cite{Penrose1,Penrose2,Penrose3}, which asserts that the existence of naked singularity is unphysical; i.e., given any regular and generic initial data, the solutions of Einstein's equation will not contain any naked singularity. The hypothesis implies that the process of gravitational collapse will always lead to a singularity hidden inside an event horizon. The exact mathematical formulation of CCH is very involved and requires many technical concepts; in particular an understanding of the global properties of the solutions of Einstein's equations \cite{Choquet-Bruhat:2009xil}. For an extensive review of the gravitational collapse and formation of naked singularities, refer to \cite{Wald, Harada, Joshi}. \\

In the absence of a general proof, an alternative approach to study CCH could be to find viable counterexamples. This was the motivation behind the classic work by Wald \cite{Wald1}, where he analyzed the possibility of overcharging an extremal Kerr-Newman black hole by throwing in test particles to create a naked singularity. If possible, this will constitute an example of the formation of a naked singularity from regular initial data using a physical mechanism. Interestingly, the physics of general relativity prohibits such a process, the test particles with sufficient charge and angular momentum, which can overcharge the black hole get repelled and does not enter the event horizon. Similar analysis has been done on various scenarios, and it is always found that the formation of a naked singularity by overcharging is not possible \cite{Semiz1,Semiz2,Cardoso,Toth,Gao,Natario,Fairoos,Jana,Akash,Vega}. On the other hand, Hubeny has shown that if the initial configuration is near extremal, it is possible to overcharge a Reissner-Nordstrom (RN) black hole using test particle absorption \cite{Hubeny}. Similar results are obtained for near extremal rotating black holes where it is shown that over-spinning can be achieved by adding either orbital or spin angular momentum using a test body \cite{Jacobson:2009kt, Dadhich}. This indicates a possible violation of the CCH and formation of the naked singularity by a physical process. However, it is also argued that the process does not take into account the back reaction effects, and later careful analysis indeed shows that appropriate consideration of the back reaction effects remove the possibility of the formation of the naked singularity \cite{Tanaka}.\\

All these analyses are performed in the context of the general relativity. But, perturbative non-renormalizability of GR strongly suggests that the theory may make sense only as an effective theory with higher curvature correction terms. The question of the validity of the CCH depends on the behavior of the theory at the high curvature regime, and these correction terms will be relevant at the same regime. So, understanding the problem of overcharging black holes in higher curvature gravity is essential to comprehend the full status of CCH. In this regard, we focus on Einstein-Gauss-Bonnet (EGB) gravity, which is the unique higher curvature extension of GR up to $D\leq 5$ with field equation not more than second order in time. The theory has well defined initial value formalism and free from perturbative ghosts in any dimension \cite{Zwiebach:1985uq}. As a result, the EGB gravity is a well-motivated model to study the effect of higher curvature terms on Wald-Hubeny type experiments.  \\

We study the overcharging problem for charged black hole solutions of Einstein-Gauss-Bonnet gravity (EGB) in general $D$ dimensions. These solutions are the generalization of Reissner-Nordstrom black holes in GR. We show that as in the case of GR, it is not possible to overcharge the extremal charged black holes in Einstein-Gauss-Bonnet gravity in any dimension. In a general $D$ spacetime dimensions, the algebraic equation for the location of the horizon of an EGB black hole is a $2(D-3)$-th degree polynomial equation. The criteria of the existence of real positive roots of such an equation are not readily available, and we need to use alternative novel physical reasoning to find the criterion of overcharging. As we will see, such a feature is unique to the higher curvature terms. Our result indicates that the principle of cosmic censorship may be valid beyond general relativity and to well motivated higher curvature theories. We also study near extremal charged black holes and show the possibility of overcharging using test particle absorption.

\section*{Charged black holes in EGB gravity}
We start with the Lagrangian of Einstein-Gauss-Bonnet gravity, ${\cal L} =  R + \hat{\alpha} \left( R^2 - 4 R_{ab} R^{ab} + R_{abcd} R^{abcd} \right)$. The spherically symmetric charged black hole solutions of this theory in $D$ spacetime dimensions is of the form \cite{Boulware:1985wk, Wiltshire1,Wiltshire2},
\begin{equation}
ds^2=-f(r)dt^2+\frac{1}{f(r)}dr^2+r^2d\Omega_{D-2}^2,
\end{equation}
with,\begin{equation}
f(r)=1+\frac{r^2}{4\alpha}\Bigg[1-\sqrt{1+\frac{16\alpha M}{r^{D-1}}-\frac{8\alpha Q^2}{r^{2D-4}}}\ \Bigg],
\end{equation}
where, the constant $\alpha$ is related to the Gauss-Bonnet coupling constant $\hat{\alpha}$ as, $\alpha =  (D-3)(D-4) \hat{\alpha} / 2$. We denote the line element by $ d\Omega_{D-2}^2 $ for the unit $(D-2)$ sphere of area $\mathcal{A}_{D-2} $. The only non-zero component of the electromagnetic vector potential has the form,
\begin{equation}
A_t(r)=-\frac{Q}{r^{D-3}} \nonumber.
\end{equation}
Here, $M$ and $Q$ are related to the black hole ADM mass ($\hat{M}$) and charge ($\hat{Q}$) as,
\begin{gather}
M = \frac{8 \pi}{(D-2)\mathcal{A}_{(D-2)}} \hat{M},\\ \nonumber
Q^2 = \frac{2(D-3)}{(D-2)} \hat{Q}^2.
\end{gather}
We will refer $M$ and $Q$ as the mass and the charge of the black hole respectively. The location of the event horizon is obtained from the zeroes of the function $f(r)$. Let the solution has a horizon at $r=r_h$ where $f(r_h)=0$, which gives,
\begin{equation}\label{horizon}
r_h^{2(D-3)}+2\alpha\, r_h^{2(D-4)}-2Mr_h^{D-3}+Q^2=0.
\end{equation}
This equation has at most two real positive roots by Descartes' rule of signs, namely $r_h=r_{\pm}$, where plus and minus signs denote the largest and the lowest positive roots respectively. Now, suppose for some combination of mass and charge, the black hole becomes extremal at $r=u$ at which the surface gravity vanishes and therefore $f(u)=f'(u)=0$. This implies,

\begin{equation}\label{extr_horizon}
u^{D-3}+\frac{2\alpha (D-4)}{D-3} u^{D-5}-M=0.
\end{equation}
This equation determines the location of the horizon of the extremal black hole.
\section*{Overcharging extremal \& near extremal black holes in EGB gravity}
We consider a test particle of mass/energy $ E$ and charge $q$ in the background described above. When the particle enters the horizon, the new mass and charge of the black hole become $M + E$ and $Q + q$. We want to choose the particle parameters such that in the new configuration, the Eq. (\ref{horizon}) has no real positive root; the horizon does not exist anymore, and what is left is a naked singularity. In the case of GR i.e., when $\alpha = 0$, Eq. (\ref{horizon}) can be regarded as a quadratic equation in $r_h^{D-3}$ and we can easily obtain an explicit condition on mass and charge so that there is no real positive root as $Q^2 > M^2$ \cite{Revelar:2017sem}. But, in the presence of the higher curvature terms, the situation is different. Eq. (\ref{horizon}) is a higher degree polynomial equation and Abel-Ruffini theorem says that there is no solution in radicals to general polynomial equations of degree five or higher with arbitrary coefficients. As a result, the criteria for having real positive roots is not straightforward. The only exception is the special case of $D=5$, for which the Eq. (\ref{horizon}) is again exactly solvable. In the absence of any such explicit solution, we need to use physical arguments to obtain a criterion for the formation of the naked singularity. To obtain such a condition, let us first discuss some properties of the metric function for different cases. First of all, asymptotic flatness implies $f(r) \rightarrow 1$ as $r \rightarrow \infty$. Then, there are the following possibilities:\\
\\(1)\ For a non-extremal black hole, $f(r)=0$ equation gives us two real positive roots ($r_{\pm}$, with $r_+ > r_-$) and $f'(r_+)>0$, since $f(r)$ has to reach the value unity as $r \rightarrow \infty$. These are the necessary and sufficient conditions for having a non-extremal black hole.\\ 
\\
\\(2)\ For an extremal black hole, $f(r)=0$ equation gives us a doubly degenerate positive real root at $r=u$ and $f'(u)=0$. This is the necessary and sufficient condition for having an extremal black hole.\\
\\(3)\ For a naked singularity, we should have no real root of the horizon equation $f(r)=0$. Since the space-time we are considering is asymptotically flat, we must have $f(r) > 0$ always. This is the necessary and sufficient condition for having a naked singularity.\\

We notice that Eq. (\ref{extr_horizon}) must have a real positive root for $D\geq 5 $. We can write this root as $u^{D-3}=M+K(\alpha,M,D)$ where, the quantity $K$ is defined as $K(\alpha,M,D)=-2\alpha (D-4)\, u^{D-5}/(D-3)$ where, $u$ is to be understood as a function of $\alpha$, $M$ and $D$. Note that, we do not have an explicit solution, but the Descartes' rule of signs guarantees the existence of such a real positive root. \\
The initial configuration is parameterized by mass M and charge Q. Now we allow a test particle of mass/energy $E$ and charge $q$ to enter the event horizon. The new system has a mass $ \bar M = M + E$ and charge $ \bar Q = Q + q$. Creating a naked singularity is tantamount to the demand that, the new configuration has no horizon; therefore, Eq. (\ref{horizon}) has no real positive root with black hole parameters as $\bar M$ and $\bar Q$. We will obtain a general condition for which the final configuration achieved by throwing the test charge of parameters (E, q) is a naked singularity, i.e., $f(r) > 0$ in the full domain. We will cast the ultimate expression in terms of the final configuration parameters ($\bar M=M+E$, $\bar Q=Q+q$).  

For a naked singularity, the possibility $(3)$ is satisfied, and we have:

\begin{equation}\label{Gr}
r^{2(D-3)}+2\alpha\, r^{2(D-4)}-2\bar M\,r^{D-3}+{\bar Q}^2 > 0, \,
\, \forall r\in [0,\infty).
\end{equation}

This polynomial is the `horizon equation' Eq. (\ref{horizon}) in disguise. The extremum value of the polynomial must be positive, as the polynomial itself is positive in the whole domain. The extremum of the above equation occurs at, say $r=\bar u>0$, which gives,
\begin{equation}\label{equbar}
{\bar u}^{D-3}+\frac{2\alpha(D-4)}{(D-3)} {\bar u}^{D-5}-\bar M=0\ ,
\end{equation}
where, ${\bar u}^{D-3}=\bar M+\bar K(\alpha,\bar M,D)$ and,
\begin{equation}
\bar K(\alpha,\bar M,D)=-\frac{2\alpha(D-4)}{D-3}{\bar u}^{D-5}.\label{Kbar}
\end{equation}
Here, $\bar u$ on the RHS of Eq.[\ref{Kbar}] is to be understood as a function which depends on ($\alpha$, $\bar M$, D). Then, we get the necessary and sufficient condition for a naked singularity by substituting $\bar u$ in Eq. (\ref{Gr}). It gives `the overcharging condition' irrespective of any initial condition on M and Q,
\begin{equation}\label{overD}
{\bar Q}^2>({\bar M}^2-{\bar K}^2)-2\alpha({\bar M}+{\bar K})^{\frac{2(D-4)}{(D-3)}}\ .
\end{equation}
By a similar approach, one can show that the necessary and sufficient condition for the extremality can be written as:
\begin{equation}\label{extremeD}
{Q}^2=({M}^2-{K}^2)-2\alpha({M}+{K})^{\frac{2(D-4)}{(D-3)}}\ .
\end{equation}
The above equation is called the `extremality condition' of the black hole.  We do not have the explicit expression of the quantity $K$, but such will not be required for demonstration of the intended result. \\

Our aim is to show that if we start with an extremal black hole satisfying Eq. (\ref{extremeD}), we can not create a configuration which satisfies Eq. (\ref{overD}) by test particle absorption. Exploiting the fact that the initial configuration is extremal, Eq. (\ref{extremeD}) and Eq. (\ref{overD}) imply,
\begin{equation}
E < \frac{q Q}{M}+\frac{({\bar K}^2-K^2)}{2M}+\frac{\alpha ({\bar u}^{2(D-4)}-u^{2(D-4)})}{M}, \label{energy_over}
\end{equation}
where we have assumed $E << M$ and $q << Q$. Our next non-trivial step is to find the relation between $u$ and $\bar{u}$. We define a polynomial function,
\begin{equation}
F(r,M) = r^{D-3}+\frac{2\alpha (D-4)}{D-3} r^{D-5}-M.
\end{equation}
The extremal horizon radius $ r =u$ satisfies $F(u,M)=0$. Also, for $D \geq 5$ the radial derivative of the above function at r=u, $F'(r=u)$ is always positive. Since $F(\bar{u},M) = E > 0$, we have $\bar{u} > u$. This immediately tells us $\bar{K} \leq K$. We also realize that $u$ and $\bar{u}$ are the roots of two polynomials with a small change in their corresponding coefficients ($M$ changes to $M+E$, where $E << M$). Considering the fact that the roots of a polynomial equation depend continuously on its coefficients, i.e., if we change the coefficients of a polynomial slightly, its root will also change slightly; one can infer that (($\bar{u}-u)/u <<1$). We expand $\bar{u}$ with respect to the small deviation parameter $E$ up to linear order as $\bar u=(1+ p \ E) u$ \ where p is a parameter so chosen that the above relation is correct dimensionally. On substitution, the overcharging condition in Eq.(\ref{energy_over}) becomes,
\begin{equation}
E<\frac{q\, Q}{u^{D-3}}.\label{finalc}
\end{equation}
This equation gives us the upper bound on the energy of the test charge to create a naked singularity out of an extremal black hole. Eq. (\ref{finalc}) is a remarkably simple result and exactly identical in form with the corresponding condition in GR. This prompts us to make a conjecture that the same form will be even valid for higher order Lovelock theories. \\
Next, we like to focus on the condition that the initial black hole can indeed absorb the particle. Since we are neglecting the black reaction effect, we consider the particle to be moving on a trajectory obtained from the solution of the Lorentz force equation. To ensure that the particle enters the horizon, we must have its radial velocity to be positive semi-definite until it reaches the horizon. This immediately gives a lower limit on the energy of the particle as \cite{Revelar:2017sem},
\begin{equation}
E>\frac{q\, Q}{u^{D-3}}.\label{entering}
\end{equation}
It is remarkable that the condition of entering Eq. (\ref{entering}) and condition of overcharging in Eq. (\ref{finalc}) can not be satisfied simultaneously, prohibiting the creation of a naked singularity from the initial extremal black hole. Notice that irrespective of the nontrivial higher curvature corrections, the situation is identical to GR. This seems to suggest that there could a simpler and elegant understanding of this result, which may not require the explicit form of the black hole solution. \\

After completing the extremal case, we can also study the possibility of the overcharging of a near-extremal black hole in $D$ dimensional EGB gravity. The essential details of the calculation are similar to that in general relativity. Here, the entering condition is given by,
\begin{equation}\label{enteringNE}
E>\frac{q Q}{r_+^{D-3}}\ .
\end{equation}
The above equation gives a lower limit on the energy of the test charge, such that it will fall past the horizon. Eq.[\ref{overD}] gives the overcharging condition as,
\begin{equation}\label{OVERD2}
(M+E)^2-2\alpha\, v^{2(D-4)}-K'^2+2E\, K'<(Q+q)^2\ ,
\end{equation}
where the quantity $v$ is a solution of the algebraic equation $F(r=v) = 0$ and, $K'=-\frac{2\alpha(D-4)}{D-3}{v}^{D-5}$. It is possible to manipulate this equation, and  the overcharging condition can be written as 
\begin{equation}\label{ED5}
E<\sqrt{{\bar Q}^2-\frac{2\alpha (D-5)}{(D-3)}\ v^{2(D-4)}}-v^{D-3}\ .
\end{equation}
The above equation gives an upper limit on the energy of the test charge, such that it may overcharge the black hole. Overcharging is possible only when the upper limit is greater than the lower limit. The entering condition in Eq. (\ref{enteringNE}) provides a lower limit on the energy/mass of the particle. If the particle parameters are chosen according to these constraints, the overcharging of the near-extremal black hole can be achieved. In GR, it is shown that the consideration of the back reaction effects can prohibit such a process \cite{Hubeny}. We conjecture the same will be true for the EGB gravity; these are details to be worked out. \\

\section*{Discussion}
The validity of the cosmic censorship depends on the properties of the evolution equations of the metric. In the case of general relativity, the global properties of Einstein's equation will decide if the conjecture holds. Also, we expect that higher curvature terms will modify GR at some sufficient large curvature scale. This may have a significant effect on the efficacy of cosmic censorship. Therefore, it is interesting to generalize the attempted counterexamples of CCH to higher curvature gravity. This is the motivation of our work where we consider the case of Einstein-Gauss-Bonnet gravity in $D$ dimensions and show that it is impossible to overcharge an extremal black hole. The situation is identical to general relativity, and higher curvature contributions conform with the cosmic censorship. \\
Our result also holds for the negative Gauss-Bonnet coupling $\alpha$, provided that there exists a valid black hole solution for the initial parameters, i.e., mass M and charge Q. Although there are other reasons to prefer the case $\alpha > 0$ \cite{Cheung:2016wjt}, no such constraint is required to negate the possibility of the formation of a naked singularity. \\
We use asymptotic flatness to obtain the necessary and sufficient condition for the formation of the naked singularity. It will be interesting to generalize our result to other asymptotic boundary condition, in particular for de Sitter and Anti-de Sitter. \\
Recently, it is proven that one can not create a naked singularity from an extremal Kerr-Newman black hole in general relativity, provided that the non-electromagnetic contribution to the stress-energy tensor of the matter satisfies the null energy condition \cite{Sorce_Wald}. The proof is sufficiently general and incorporates all possible self-force and back reaction effects up to the second order correction to mass. Our result suggests that such proof may also be valid for the Einstein-Gauss-Bonnet gravity. \\

\section*{Acknowledgement}
We thank Naresh Dadhich, Avirup Ghosh \& Akash K Mishra for many helpful discussions. Research of SS is supported by the Department of Science and Technology, Government of India under the SERB Matrics Grant (MTR/2017/000399). SS also thanks the hospitality of the Abdus Salam International Centre for Theoretical Physics (ICTP) where a part of this work is completed.

\end{document}